\documentclass[12pt]{article}
\usepackage{amsmath}
\usepackage{color}
\begin{document}
\begin{flushright}
KANAZAWA-15-14\\
September, 2015
\end{flushright}
\vspace*{1cm}

\begin{center} 
{\Large\bf Inflation due to a non-minimal coupling of singlet scalars
in the radiative seesaw model}
\vspace*{1cm}

{\Large Romy H. S. Budhi}~$^{*,\dagger,}$\footnote{e-mail:
~romyhanang@hep.s.kanazawa-u.ac.jp, romyhanang@ugm.ac.id}, ~
{\Large Shoichi Kashiwase}~$^{\dagger,}$\footnote{e-mail:
  ~shoichi@hep.s.kanazawa-u.ac.jp},\\
{\Large and}\\
{\Large Daijiro Suematsu}~$^{\dagger,}$\footnote{e-mail:
~suematsu@hep.s.kanazawa-u.ac.jp}
\vspace*{0.5cm}\\
$^*${\it Physics Department, Gadjah Mada University, 
Yogyakarta 55281, Indonesia}\\
$^\dagger${\it Institute for Theoretical Physics, Kanazawa University, 
Kanazawa 920-1192, Japan}
\end{center}
\vspace*{1.5cm} 

\noindent
{\Large\bf Abstract}\\
The radiative neutrino mass model with inert doublet dark matter 
is a promising model for the present experimental issues
which cannot be explained within the standard model.  
We study an extension of this model focusing on cosmological features 
brought about from the scalar sector.
Inflation due to singlet scalars with hierarchical non-minimal couplings 
with the Ricci scalar may give a favorable solution for both neutrino masses 
and baryon number asymmetry in the Universe.        
\newpage
\section{Introduction}
Recent discovery of the Higgs particle \cite{higgs} suggests that 
the framework of the standard model (SM) can describe Nature up to 
the weak scale.
On the other hand, we have several experimental results which are 
not explained within it. Representative ones are the existence of both small 
neutrino masses \cite{nexp} and dark matter (DM) \cite{uobs}, and also 
the baryon number asymmetry in the Universe \cite{baryon}. 
These require some extension of the SM.

As such an extension, we have an interesting simple model, 
in which the SM is extended by an inert doublet scalar and 
singlet fermions \cite{ma}.
It has promising features for the simultaneous explanation 
of the neutrino oscillation data \cite{nexp} and the observed 
abundance of DM \cite{uobs} through physics at TeV regions.
In fact, if these new fields are assigned odd parity of an imposed $Z_2$
symmetry, the neutrino masses are generated at one-loop level and
the lightest $Z_2$ odd particle can behave as DM.
The quantitative conditions required for their explanation and also other
phenomenological aspects have been extensively studied 
in this model and its extended models \cite{fcnc,flavor,u1,ma_type,nonth,ks}.
They show that the simultaneous explanation of these is possible 
without causing strong tension with other phenomena like lepton 
flavor violating processes as long as DM is identified with the lightest 
neutral component of the inert doublet scalar \cite{ks,ham}.\footnote{If the
lightest singlet fermion is identified with DM, strong tension
appears between the DM abundance and the lepton flavor violating
processes \cite{fcnc}. However, it could be resolved by assuming 
special flavor structure \cite{flavor} or introducing a 
new interaction \cite{u1}.} 
Moreover, in that case, the baryon number asymmetry in the Universe 
could be also successfully explained if rather mild mass degeneracy 
is assumed among the singlet fermions with masses of TeV scales \cite{ks}.

In this paper, we consider how inflation can be embedded in this
framework. 
CMB observations \cite{planck,bkp,planck15} suggest that 
the exponential expansion of the Universe should occur before 
the ordinary Big-Bang of the Universe. On the other hand, the analyses 
of them seem to have already ruled out a lot of inflation models proposed 
by now. 
Higgs inflation is a well-known example which is still alive \cite{higgsinf}.
This model is characterized by the feature such that 
Higgs potential becomes flat enough for large field regions 
if the Higgs scalar has a large non-minimal coupling with the Ricci scalar. 
We apply this idea to a radiative seesaw model extended 
by real singlet scalars \cite{ext-s}.
Although the singlet scalars are originally introduced with the aim of 
generating the neutrino masses, it could work as inflaton if they are 
supposed to have a substantial non-minimal coupling with the Ricci scalar.
In fact, such a coupling of a real singlet scalar has been studied 
as $s$-inflation in a different context \cite{sinfl}.
Following it, we focus our attention on such a non-minimal coupling of the
singlet scalars with the Ricci scalar instead of the one of Higgs scalar 
and others.  
In this case, unitarity problem which appears in the Higgs inflation 
and other general models \cite{unitarity1,unitarity2} might be escaped 
under certain conditions. 
Moreover, the singlet scalars could also play an important role in the 
generation of the baryon number asymmetry in the Universe through 
non-thermal leptogenesis. We study this issue intensively. 

The following parts of the paper are organized as follows.
In section 2, we introduce the model studied in this paper and discuss
the neutrino mass generation. 
In section 3, a possible inflation scenario in this model is discussed. 
Production of the baryon number asymmetry due to the inflaton 
decay is studied in detail in section 4. Consistency of the DM physics with
this scenario is also discussed here. 
Section 5 is devoted to the summary of the paper.

\section{An extension with real singlet scalars}
The radiative seesaw model proposed in \cite{ma} is characterized 
by a scalar quartic coupling $\lambda_5(\eta^\dagger\phi)^2$ between 
the ordinary Higgs doublet $\phi$ and the inert doublet scalar $\eta$.
Since $\eta$ and singlet fermions $N_k$ are assigned odd parity of
the $Z_2$ symmetry and all the SM contents are assigned even parity, the
Dirac neutrino mass terms are forbidden at tree level.
Neutrino masses are generated through a one-loop diagram with $N_k$ 
and $\eta$ in the internal lines. In this mass generation scenario 
at TeV scales, the above mentioned quartic coupling between $\phi$ 
and $\eta$ plays an essential role to explain the small neutrino masses. 

An extension of the model might be done by considering a 
possibility that this quartic coupling is an effective coupling 
appearing at low energy regions after integrating out heavy scalar fields 
\cite{ext-s}. 
Such a scenario could be realized by introducing $Z_2$ odd real 
singlet scalars $S_a~(a=1,2)$.\footnote{One real scalar 
is enough for the neutrino mass generation and inflation.
However, if we consider leptogenesis in the model, two real scalars 
should be introduced at least. We take this minimal version here.}  
The model is defined by a part of 
Lagrangian relevant to the new fields as 
follows,
\begin{eqnarray}
-{\cal L}&=&\sum_{\alpha,k=1}^3\left[h_{\alpha k} \bar N_k\eta^\dagger\ell_\alpha
+h_{\alpha k}^\ast\bar\ell_\alpha\eta N_k+
\frac{M_k}{2}\bar N_kN_k^c 
+\frac{M_k}{2}\bar N_k^cN_k\right] \nonumber \\
&+&m_\phi^2\phi^\dagger\phi+m_\eta^2\eta^\dagger\eta+
\lambda_1(\phi^\dagger\phi)^2+\lambda_2(\eta^\dagger\eta)^2
+\lambda_3(\phi^\dagger\phi)(\eta^\dagger\eta)  
+\lambda_4(\eta^\dagger\phi)(\phi^\dagger\eta) \nonumber \\ 
&+&\sum_{a=1,2}\left[\frac{m_{S_a}^2}{2}S_a^2
+\frac{\kappa_{1}^{(a)}}{4}S_a^4+\frac{\kappa_{2}^{(a)}}{2}
S_a^2(\phi^\dagger\phi)
+\frac{\kappa_{3}^{(a)}}{2}S_a^2(\eta^\dagger\eta)+\mu_a S_a\eta^\dagger\phi
+ \mu_a^\ast S_a\phi^\dagger\eta\right], \nonumber \\
\label{model}
\end{eqnarray}
where $\ell_\alpha$ is a left-handed doublet lepton.
We note that $\lambda_5(\eta^\dagger\phi)^2$ is allowed under 
the imposed symmetry in general. However, if we assume $\lambda_5=0$ 
in the original Lagrangian, its $\beta$-function is proportional to itself 
and then $\lambda_5=0$ is stable against the radiative correction 
as long as $\mu_a$ terms are not included in eq.~(\ref{model}). 
On the other hand, if the $\mu_a$ terms are introduced in eq.~(\ref{model})
assuming $\lambda_5=0$, the $\lambda_5$ term appears effectively 
as discussed below. Later, this point will be discussed again in relation 
to the assignment of lepton number to the new fields.

\input epsf
\begin{figure}[t]
\begin{center}
\epsfxsize=8cm
\leavevmode
\epsfbox{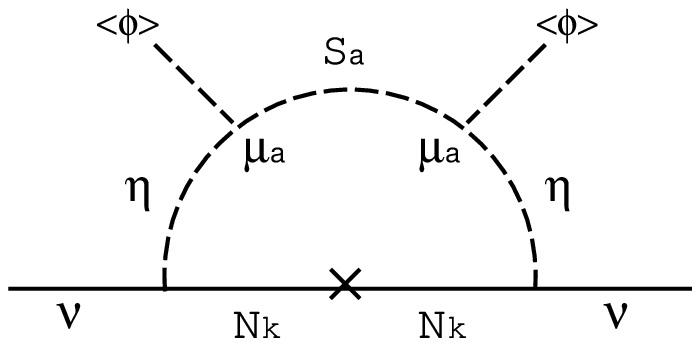}
\end{center}
\vspace*{-3mm}

{\footnotesize {\bf Fig.~1}~~The one-loop diagram which contributes 
neutrino mass generation in the present model.}  
\end{figure}

In this model, neutrino masses are generated through a one-loop diagram
which is shown in Fig.~1. They are estimated as
\begin{equation}
{\cal M}^\nu_{\alpha\beta}=\sum_{k=1}^3
\frac{h_{\alpha k}h_{\beta k}M_k\langle\phi\rangle^2}{8\pi^2}
\sum_a\mu_a^2I(M_\eta,M_k,m_{s_a}),
\label{nmass1}
\end{equation}
where $M_\eta^2=m_\eta^2+(\lambda_3+\lambda_4)\langle\phi\rangle^2$ and 
$\langle\phi\rangle=174$~GeV. The function $I$ is defined by
\begin{eqnarray}
I(m_a,m_b,m_c)&=&\frac{(m_a^4-m_b^2m_c^2)\ln m_a^2}{(m_b^2-m_a^2)^2(m_c^2-m_a^2)^2}
+\frac{m_b^2\ln m_b^2}{(m_c^2-m_b^2)(m_a^2-m_b^2)^2} \nonumber \\
&+&\frac{m_c^2\ln m_c^2}{(m_b^2-m_c^2)(m_a^2-m_c^2)^2}
-\frac{1}{(m_b^2-m_a^2)(m_c^2-m_a^2)}.
\label{nmass2}
\end{eqnarray}
If we suppose $m_{S_a}\gg M_\eta,~M_k$, this formula can be approximated as
\begin{equation}
{\cal M}^\nu_{\alpha\beta}=\sum_a\frac{\mu_a^2}{m_{S_a}^2}\sum_{k=1}^3
\frac{h_{\alpha k}h_{\beta k}\langle\phi\rangle^2}{8\pi^2}
\frac{M_k}{M_\eta^2-M_k^2}
\left[\frac{M_k^2}{M_\eta^2-M_k^2}\ln\frac{M_k^2}{M_\eta^2}+1\right].
\label{nmass3}
\end{equation}
This is equivalent to the neutrino mass formula in the original model
if $\sum_a\frac{\mu_a^2}{m_{S_a}^2}$ is identified with the quartic coupling 
constant $\lambda_5$ \cite{ext-s}. 
This correspondence could be directly confirmed in an effective model 
at energy regions smaller than $m_{S_a}$, which can be derived 
by integrating out $S_a$. In fact, 
since the equation of motion for $S_a$ could be approximated as 
$S_a\simeq -\frac{1}{m_{S_a}^2}(\mu_a\eta^\dagger\phi+\mu_a^\ast\phi^\dagger\eta)$,
the required terms are derived by using it as
\begin{equation}
-\frac{1}{2}\sum_a\left[\frac{\mu_a^2}{m_{S_a}^2}(\eta^\dagger\phi)^2
+\frac{\mu_a^{\ast 2}}{m_{S_a}^2}(\phi^\dagger\eta)^2\right].
\end{equation} 
Origin of the smallness of $|\lambda_5|$, which is a key to explain 
the small neutrino masses in the original model, is translated 
to the hierarchy problem between $\mu_a$ and $m_{S_a}$ in this scenario.
We cannot answer the origin of this hierarchy at the present stage and 
we have to leave it for a complete theory at high energy regions.

For the later study, we show an example of flavor structure of 
the neutrino Yukawa couplings which can explain 
every neutrino oscillation data in the normal hierarchy case.
Here, we follow the procedure given in \cite{ks}.  
For this purpose, we assume that the neutrino mass matrix (\ref{nmass3}) 
takes the following simple form as
\begin{equation}
{\cal M}^\nu=
\left(
\begin{array}{ccc}
0 & 0 & 0\\ 0 & 1 & q_1 \\ 0 & q_1 & q_1^2 \\ 
\end{array}\right)(h_1^2\Lambda_1+ h_2^2\Lambda_2)\frac{\mu_2^2}{m_{S_2}^2}+
\left(
\begin{array}{ccc}
1 & q_2 & -q_3\\ q_2 & q_2^2 & -q_2q_3 \\ -q_3 & -q_2q_3 & q_3^2 \\ 
\end{array}\right)h_3^2\Lambda_3\frac{\mu_2^2}{m_{S_2}^2},
\label{nmass4}
\end{equation} 
where $\frac{|\mu_1|^2}{m_{S_1}^2}\ll\frac{|\mu_2|^2}{m_{S_2}^2}$ is assumed\footnote{As explained in the later discussion, this assumption is
adopted in connection with leptogenesis.}
and $\Lambda_k$ is represented as
\begin{equation}
\Lambda_k=\frac{\langle\phi\rangle^2}{8\pi^2~{\rm GeV}}
\frac{\frac{1{\rm GeV}}{M_k}}{\frac{M_\eta^2}{M_k^2}-1}
\left[1+\frac{M_k^2}{M_\eta^2-M_k^2}\ln\frac{M_k^2}{M_\eta^2}\right]
\equiv  \frac{\langle\phi\rangle^2}{8\pi^2~{\rm GeV}}\tilde\Lambda_k.
\end{equation}

If we put $q_{1,2,3}=1$ in eq.~(\ref{nmass4}), the Pontecorvo-Maki-Nakagawa-Sakata (PMNS) mixing matrix is 
found to reduce to the tri-bimaximal form 
\begin{equation}
U_{PMNS}=\left(\begin{array}{ccc}
\frac{2}{\sqrt 6} & \frac{1}{\sqrt 3} & 0\\
 \frac{-1}{\sqrt 6} & \frac{1}{\sqrt 3} & \frac{1}{\sqrt 2}\\
\frac{1}{\sqrt 6} & \frac{-1}{\sqrt 3} & \frac{1}{\sqrt 2}\\
\end{array}\right)
\left(\begin{array}{ccc}
1 &0 & 0\\
0 & e^{i\alpha_1} & 0 \\
0 & 0 & e^{i\alpha_2} \\
\end{array}\right),
\label{mns} 
\end{equation}
where Majorana phases $\alpha_{1,2}$ are determined by the phases 
$h_i$ and $\mu_a$. If we put $\varphi_i={\rm arg}(h_i)$ and 
$\varphi_{\mu_a}={\rm arg}(\mu_a)$, they are expressed as
\begin{equation}
\alpha_1=\varphi_3 +\varphi_{\mu_2}, \qquad
\alpha_2=\varphi_2+\varphi_{\mu_2},
\end{equation}
where $|h_1|$ is taken as a negligibly small value compared with 
others, for simplicity.\footnote{The model is equivalent 
to the one with two singlet fermions in this case. 
It should be noted that the neutrino oscillation data could be 
explained as long as only two singlet fermions 
are introduced. We use this setting throughout the following study.} 
Since one of mass eigenvalues is always zero in this flavor structure, 
we find that the mass eigenvalues should satisfy
$|h_2|^2\Lambda_2\frac{|\mu_2|^2}{m_{S_2}^2} \simeq 
\frac{\sqrt{\Delta m_{\rm atm}^2}}{2}$ and
$|h_3|^2\Lambda_3\frac{|\mu_2|^2}{m_{S_2}^2}
\simeq \frac{\sqrt{\Delta m_{\rm sol}^2}}{3}$,
where $\Delta m^2_{\rm atm}$ and $\Delta m^2_{\rm sol}$ stand for 
the squared mass differences required by the neutrino oscillation 
analysis for both atmospheric and solar neutrinos \cite{nexp,pdg}.

\begin{figure}[t]
\begin{center}
\epsfxsize=8cm
\leavevmode
\epsfbox{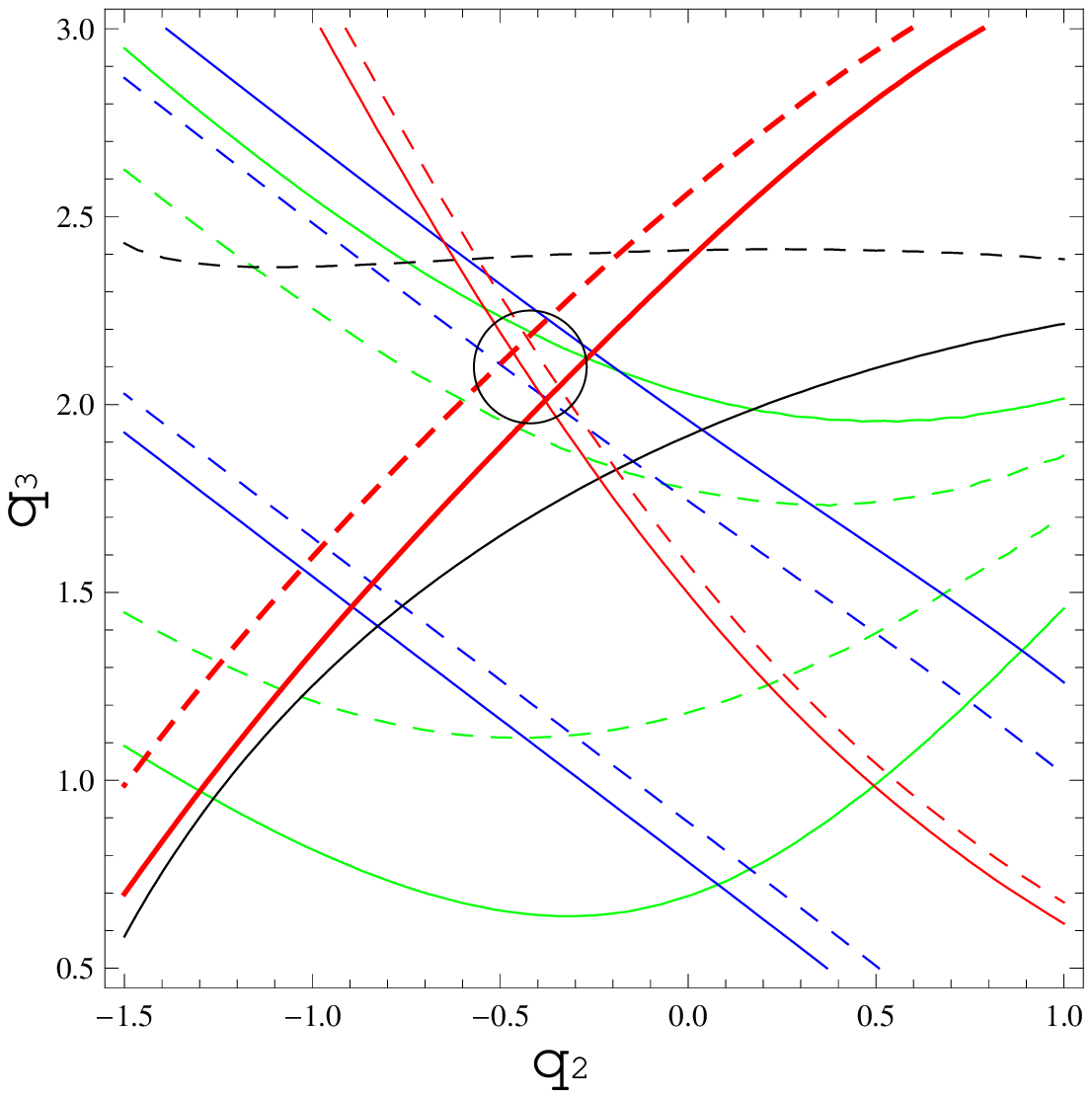}
\end{center}
\vspace*{-3mm}

{\footnotesize {\bf Fig.~2}~~A region in the $(q_2,q_3)$ plane
allowed by the neutrino oscillation data, which is included 
inside a circle drawn by the black solid line.
Other parameters are fixed to satisfy the condition (\ref{parameters}).
Each contour represents 
2$\sigma$ boundary values of neutrino oscillation parameters \cite{pdg},
that is, $|\Delta m^2_{32}|$ (thick red solid and dashed lines), 
$\Delta m_{21}^2$ (thin red solid and dashed lines), 
$\sin^22\theta_{23}$ (green solid and dashed lines), 
$\sin^22\theta_{12}$ (blue solid and dashed lines) 
and $\sin^22\theta_{13}$ (black solid and dashed lines).}  
\end{figure}

Since we now know that $\sin\theta_{13}$ takes a non-zero value,
we have to consider a flavor structure deviated from $q_{1,2,3}=1$. 
For that purpose, we determine values of $q_{1,2,3}$, 
$h_2^2\tilde\Lambda_2\frac{|\mu_2|^2}{m_{S_2}^2}$ and 
$h_3^2\tilde\Lambda_3\frac{|\mu_2|^2}{m_{S_2}^2}$ 
to realize all the squared mass differences and the
mixing angles required by the neutrino oscillation data
through diagonalizing the matrix (\ref{nmass4}) numerically. 
This analysis shows that the neutrino 
oscillation parameters can be in the 2$\sigma$ range of the experimental
data if the values of $(q_2,q_3)$ are contained in the region surrounded by 
a circle in Fig.~2. In this figure, the remaining parameters are fixed 
so as to satisfy
\begin{equation}
q_1=0.77, \quad  
h_2^2\tilde\Lambda_2\frac{|\mu_2|^2}{m_{S_2}^2}= 6.03\times 10^{-14}, \quad
h_3^2\tilde\Lambda_3\frac{|\mu_2|^2}{m_{S_2}^2}= 1.02\times 10^{-14}.
\label{parameters}
\end{equation} 
As long as the model parameters $\frac{\mu_2}{m_{S_2}}$, $M_\eta$, 
$h_{2,3}$ and $M_{2,3}$ are varied by keeping these conditions, 
the neutrino oscillation constraints are automatically fulfilled. 
In Table~1, typical examples obtained by this simple procedure are shown.
They include examples such that the masses of the singlet fermions $M_k$ are
largely different in the case $M_\eta=1$~TeV, in which they are of 
$O(10^4)$~GeV and $O(10^8)$~GeV in the cases (a) and (b), respectively.

\begin{figure}[t]
\begin{center}
\small
\begin{tabular}{c|ccccccc}\hline
$M_\eta$(GeV)&$\frac{|\mu_1|}{m_{S_1}}$ 
& $\frac{|\mu_2|}{m_{S_2}}$ & 
$h_2$ & $h_3$ & $M_2$(GeV) & $M_3$(GeV) & $|Y_B|$  \\ \hline\hline 
$10^3$\quad (a) & $2\cdot 10^{-5}$ &$ 10^{-3}$ & 
$1.12\cdot 10^{-2}$ & $5.13\cdot 10^{-3}$&
$5.30\cdot 10^3$& $9.00\cdot 10^3$  & $1.3\cdot 10^{-9}$ \\ 
\hspace*{9mm} (b) & $2\cdot 10^{-6}$ & $8\cdot 10^{-2}$ & 
$8.40\cdot 10^{-3}$& $3.85\cdot 10^{-3}$& 
$1.73\cdot 10^8$ & $2.19\cdot 10^8$ & $3.5\cdot 10^{-10}$  \\ \hline
600 & $2\cdot 10^{-5}$  & $10^{-3}$ & $9.62\cdot 10^{-3}$ &
$4.54\cdot 10^{-3}$ & $5.30\cdot 10^3$& $9.00\cdot 10^3$ & $5.1\cdot 10^{-10}$\\
$3\cdot 10^3$ & $2\cdot 10^{-5}$  & $10^{-3}$ & $2.15\cdot 10^{-3}$ &
$8.94\cdot 10^{-3}$ & $2.67\cdot 10^4$& $2.77\cdot 10^4$ & $3.4\cdot 10^{-9}$ 
\\ \hline\hline
\end{tabular}
\end{center}
\vspace*{2mm}

{\footnotesize {\bf Table 1}~~Examples of the model parameters 
which satisfy the latter two conditions in eq.~(\ref{parameters}).
If we fix a point in a circle of Fig.~2, all the neutrino oscillation 
data can be explained at 2$\sigma$ level, consistently. 
In all cases, $m_{S_1}=10^9$~GeV and $\frac{m_{S_2}}{m_{S_1}}=1.1$ are assumed.}
\end{figure} 

\normalsize
If $S_a$ does not play any other role than the neutrino 
mass generation, this extension may not be so appealing. 
However, we can find that the introduction of $S_a$ could 
add favorable features as an inflation model to the radiative 
seesaw model.\footnote{We have proposed another inflation scenario 
in the similar context based on somewhat different motivation 
in \cite{bks}. The inflaton potential assumed there differs from 
the present one. As a result, the predicted values 
for the spectral index and the tensor-to-scalar ratio take 
distinct values from the present ones.} 
Recent Planck data suggest that the Higgs inflation scenario 
could be one of the favored inflation models.  
However, if multi-component scalars like the Higgs doublet scalar 
are supposed to play a role of inflaton, the model could be suffered 
from the unitarity problem \cite{unitarity1,unitarity2}. 
Since unitarity could be violated in the scattering amplitudes
among scalars with non-minimal couplings to the Ricci scalar 
at a lower scale compared with the inflation scale, 
new physics required for unitarity restoration could jeopardize 
the flatness of inflaton potential at the inflation scale.
The situation can be changed in a real singlet inflaton as 
discussed in \cite{unitarity2}.   
In the following part, we consider that only the singlet scalars
among scalars in the model have non-negligible 
non-minimal couplings with the Ricci scalar. 

\section{Inflation due to the non-minimal coupling}
It has been known that a scalar field coupled with the Ricci scalar
can bring about the exponential expansion of the Universe \cite{nonm-inf}.
Using this idea to the SM, Higgs inflation has been proposed 
in \cite{higgsinf} as a scenario with a realistic inflaton candidate.
After that, the scenario has been studied from various view points 
\cite{h-inf1}. 
We apply this idea to the singlet scalars in this model 
but not to the Higgs doublet or 
the inert doublet.\footnote{The study of Higgs inflation 
in the inert doublet model can be found in \cite{inert-inf}.
Although the present inflation scenario and its prediction are 
essentially the same as \cite{higgsinf,sinfl}, we note that the 
inflaton is shown to play crucial roles in the neutrino mass 
generation and the leptogenesis in this model.}
The action relevant to the present inflation scenario is given
in the Jordan frame as
\begin{equation}
S_J = \int d^4x\sqrt{-g} \left[\frac{1}{2}M_{\rm pl}^2R + 
\sum_a\frac{1}{2}\xi_a S_a^2 R
+\sum_a\frac{1}{2}\partial^\mu S_a\partial_\mu S_a - V(S_a) \right],
\label{inflag}
\end{equation}
where $M_{\rm pl}$ is the reduced Planck mass and $V(S_a)$ stands for 
the corresponding part of the $S_a$ potential in eq.~(\ref{model}). 
We note that only the singlet scalars are assumed 
to have non-minimal coupling with the Ricci scalar.

We take $S_1$ as inflaton and other scalars are assumed to have 
much smaller values than $S_1$ during the inflation.
In that case, $V(S_a)$ can be approximately expressed as 
$V(S_a)\simeq\frac{\kappa_1^{(1)}}{4}S_1^4$ for a sufficiently large 
value of $S_1$, where the coupling $\kappa_i^{(1)}$ of inflaton 
is abbreviated as $\kappa_i$ and $\kappa_1S_1^2 \gg m_{S_1}^2$ is
supposed implicitly.
In order to derive the corresponding action to eq.~(\ref{inflag}) 
in the Einstein frame, 
we use the conformal transformation \cite{higgsinf,nonm-inf}
\begin{equation}
g=\Omega^2g_E,  \qquad  \Omega^2= 1+\frac{\sum_a\xi_a S_a^2}{M_{\rm pl}^2}.
\end{equation}
As a result of this transformation, we find that it is written as    
\begin{equation}
S_E=\int d^4x\sqrt{-g_E}\left[ \frac{1}{2}M_{\rm pl}^2R_E+
\frac{1}{2\Omega^4}\sum_{a,b=1,2}\left(\delta_{ab}
+ \frac{\xi_a\delta_{ab}S_a^2+6\xi_a\xi_bS_aS_b}{M_{\rm pl}^2}\right) 
\partial^\mu S_a\partial_\mu S_b
- \frac{1}{\Omega^4}V(S_a)\right].
\label{elag}
\end{equation}
An important feature is the appearance of the mixing
between $\partial_\mu S_a$ and $\partial_\mu S_b$ in the second term. 
We will discuss it later. 

We consider a case in which only one real 
scalar $S_1$ has the non-minimal coupling with the Ricci scalar at first.
In that case, a canonically normalized field $\chi$ can be 
introduced as 
\begin{equation}
\frac{d\chi}{dS_1}
=\frac{\left[1 +(\xi_1+6\xi_1^2)\frac{S_1^2}{M_{\rm pl}^2}\right]^{1/2}}
{1+\frac{\xi_1S_1^2}{M_{\rm pl}^2}}.
\label{chi}
\end{equation}
The potential $\frac{1}{\Omega^4}V(S_1)$ can be expressed by using 
this $\chi$.
It is easily seen that the new field $\chi$ coincides 
with $S_1$ at the regions where $S_1\ll \frac{M_{\rm pl}}{\sqrt\xi_1}$ 
is satisfied.
On the other hand, if $S_1$ takes a large value such as 
$S_1\gg \frac{M_{\rm pl}}{\sqrt{\xi_1}}$,
$S_1$ and $\chi$ are found from eq.~(\ref{chi}) to be related as  
$S_1\propto\exp\left(\frac{\chi}{\sqrt{6+\frac{1}{\xi_1}}M_{\rm pl}}\right)$. 
The potential at this region is found to be almost constant as
\begin{equation}
V_E^{(1)}=\frac{\kappa_1 S_1^4}{4\left(1+\frac{\xi_1 S_1^2}{M_{\rm pl}^2}
\right)^2}\simeq \frac{\kappa_1 M_{\rm pl}^4}{4\xi_1^2}.
\label{infpot}
\end{equation}   
This suggests that $\chi$ could play a role of slow-rolling inflaton 
in this region.

The number of e-foldings induced by the potential $V_E^{(1)}$ can be 
estimated as 
\begin{equation}
N=\frac{1}{M_{\rm pl}^2}\int_{\chi_{\rm end}}^\chi d\chi 
~\frac{V_E^{(1)}}{V_E^{(1)\prime}}
\simeq \frac{3}{4}\frac{S_1^2-S_{1,{\rm end}}^2}{M_{\rm pl}^2/\xi_1}, 
\label{efold}
\end{equation}
where $V_E^{(1)\prime}=\frac{dV_E^{(1)}}{d\chi}$ and eq.~(\ref{chi}) is used.
Slow roll parameters derived from this potential are summarized as
\cite{revinf}
\begin{equation}
\varepsilon=\frac{1}{M_{\rm pl}^2}\left(\frac{V_E^{(1)\prime}}{V_E^{(1)}}\right)^2=
\frac{4M_{\rm pl}^4}{3\xi_1^2S_1^4},   \qquad
\eta=M_{\rm pl}^2\left(\frac{V_E^{(1)\prime\prime}}{V_E^{(1)}}\right)=
-\frac{4M_{\rm pl}^2}{3\xi_1 S_1^2}.  
\label{slow}
\end{equation}
Since the inflation is considered to end at $\varepsilon\simeq 1$, we have 
$S_{1,{\rm end}}^2\simeq \sqrt{\frac{4}{3}}\frac{M_{\rm pl}^2}{\xi_1}$, 
which suggests that $S_{1,{\rm end}}$ could be neglected in eq.~(\ref{efold}).
Thus, the slow roll parameters are found to be expressed as
$\varepsilon \simeq \frac{3}{4}N^{-2}$ and $\eta\simeq -N^{-1}$ 
by using the $e$-foldings $N$ only.

The spectrum of density perturbation predicted by the inflation 
is expressed as
\begin{equation}
{\cal P}(k)=A_s\left(\frac{k}{k_\ast}\right)^{n_s-1},  \qquad
A_s=\frac{V_E^{(1)}}{24\pi^2M_{\rm pl}^4\varepsilon}\Big|_{k_\ast}. 
\label{power}
\end{equation}
If we use $A_s=(2.445\pm0.096)\times 10^{-9}$ 
at $k_\ast=0.002~{\rm Mpc}^{-1}$ \cite{planck}, we find that the relation 
$\kappa_1\simeq 10^{-6}\xi_1^2N^{-2}$ should be satisfied at the 
horizon exit time of the scale $k_\ast$.
The spectral index $n_s$ and the tensor-to-scalar ratio $r$
are represented by using the slow-roll parameters as \cite{revinf} 
\begin{equation}
n_s=1-6\varepsilon+2\eta, \qquad  r=16\varepsilon.
\end{equation} 
Using eq.~(\ref{slow}) to these formulas, they are found to be 
determined only by the $e$-foldings $N$ such as $n_s\sim 0.968$ 
and $r\sim 3.0\times 10^{-3}$ for $N=60$.
These values coincide with the ones estimated from the Planck data well.
Although all these results are the same as the ones found 
in the Higgs inflation, the quartic coupling $\kappa_1$ is a free 
parameter in this model.
It is completely different from the Higgs inflation where 
the corresponding quartic coupling is constrained by the Higgs mass 126 GeV.
As a result, we cannot relate weak scale physics to the inflation 
through the observational data of the Universe in this model.
On the contrary, this fact allows that $\xi_1$ takes a much smaller value 
compared with the one of the usual Higgs inflation.
For example, $\xi_1=O(10^2)$ can realize both $N=60$ and the observed value of 
$A_s$ if a very small value such as $O(10^{-6})$ is assumed for $\kappa_1$.
However, as found from the expression of the slow-roll parameters which depend 
only on $N$, the predicted values for $n_s$ and $r$ are the same as those of 
the Higgs inflation.  

Next, we consider the model with two real scalars, which corresponds to
the one discussed in the previous section.
The situation could largely change if multi-scalars have couplings 
with the Ricci scalar. 
In general, the mixing in eq.~(\ref{elag}) cannot be resolved by any 
field redefinition and then it is difficult to find canonically 
normalized basis for them. 
An exceptional situation for this could be found for hierarchical couplings 
such as $\xi_1\gg 1$ and $\xi_1\xi_2\ll 1$. 
This condition can be freely imposed 
on the present model since $S_1$ and $S_2$ are not related by any symmetry.
In that case, the model could behave as a single 
real field model \cite{unitarity2}.
We can introduce a canonically normalized field $\chi$ for $S_1$ 
in the same way as eq.~(\ref{chi}).
On the other hand, the $S_2$ relevant terms in eq.~(\ref{elag}) are
strongly suppressed as long as $\xi_1S_1^2>M_{\rm pl}^2$ is satisfied.
In the potential $V$, the $S_2$ relevant part can be given as
\begin{equation}
V_E^{(2)}=\frac{\kappa_1^{(2)} S_2^4}
{2\left(1+\frac{\xi_1 S_1^2}{M_{\rm pl}^2}\right)^2}
\simeq \frac{\kappa_1^{(2)}M_{\rm pl}^4}{2\xi_1^2}\left(\frac{S_2}{S_1}\right)^4 
 \ll V_E^{(1)}.
\end{equation}  
This means that only the $\chi$ could play a role of slow-roll inflaton 
also in this case.

In addition, under the condition $\xi_1\xi_2\ll 1$, 
the scale of unitarity violation could be 
comparable to the inflation scale $\frac{M_{\rm pl}}{\sqrt{\xi_1}}$ 
\cite{unitarity2}. 
The unitarity violating scattering induced by the mixing 
part in the second term in eq.~(\ref{elag}) could give 
the strongest constraint. 
A simple power counting for the scattering amplitude between 
$S_1$ and $S_2$ suggests that the unitarity violating scale 
is given by $\Lambda=\frac{M_{\rm pl}}{\sqrt{\xi_1\xi_2}}$.
However, since the condition $\sqrt{\xi_1\xi_2}<\sqrt{\xi_1}$ is satisfied,
the unitarity violating scale $\Lambda$ could be comparable to or larger 
than the inflation scale $\frac{M_{\rm pl}}{\sqrt{\xi_1}}$.  
Since $S_1$ and $S_2$ are independent fields in the present model, 
$\xi_2\ll 1$ is possible even if we assume a suitable value of $\xi_1$ 
for inflation. 
Other possible unitarity violation induced by other parts such as $V_E$
might be studied through the analysis by taking account of the background 
field dependence. It suggests that the unitarity violation
scale is comparable to the inflation scale or larger than that. 
Thus, the flatness of the present inflaton potential is reliable 
throughout the inflation period.
Any physics which remedies the unitarity violation does not affect
the present inflation scenario.

\section{Non-thermal leptogenesis and dark matter}
\subsection{Leptogenesis}
Reheating after inflation is another important problem 
for this inflation scenario to be realistic.
If we impose the existence of sufficient thermal relics of the inert
doublet DM $\eta_R^0$ in the present Universe, reheating temperature 
should be higher than its mass $M_{\eta_R^0}$ at least, which is supposed 
to be of $O(1)$~TeV in the present study.
Since the allowed decay mode for the inflaton $S_1$ is limited to 
$S_1\rightarrow \eta^\dagger\phi,~\phi^\dagger\eta$, 
the decay width of $S_1$ could be estimated as
$\Gamma_{S_1}=\frac{1}{8\pi}\frac{|\mu_1|^2}{m_{S_1}}$. 
Applying instantaneous thermalization approximation to this process, 
the reheating temperature could be estimated from the condition 
$H\simeq \Gamma_{S_1}$ as
\begin{equation}
T_R\simeq 1.74g_\ast^{-1/4}(\Gamma_{S_1} M_{\rm pl})^{1/2}
\simeq 1.6\times 10^{12}\left(\frac{|\mu_1|}{m_{S_1}}\right)^{1/2}
\left(\frac{|\mu_1|}{10^8~{\rm GeV}}\right)^{1/2}~{\rm GeV},
\label{tr}
\end{equation}
where $g_\ast=116$ is used.
If $T_R>M_{\eta^0_R}$ is satisfied and $\eta_R^0$ exists in the thermal bath,
the observed DM abundance could be explained as the relic abundance 
of this thermal $\eta_R^0$. The relic abundance is discussed in 
the next subsection. 

Several leptogenesis scenarios may be considered in this reheating 
processe depending on the lepton number assignment for new ingredients. 
First, we consider an ordinary lepton number assignment 
to the new fields $\eta$ and $N_k$ such that $L(\eta)=0$ and $L(N_k)=1$.
In this case, if the reheating temperature is higher than a certain bound 
required for the heavy singlet fermion masses,\footnote{As discussed 
in \cite{ks}, this bound for the singlet fermion masses could be relaxed 
in the radiative seesaw model in comparison with the famous 
Davidson-Ibarra bound in the usual seesaw model \cite{di}.} 
the usual thermal leptogenesis could work.
On the other hand, if the reheating temperature is not so high but
high enough to thermalize the singlet fermions 
with masses of $O(1)$ TeV for example, the sufficient baryon number 
asymmetry could be generated through the resonant 
leptogenesis \cite{resonant,resonant1}
as long as the masses of the singlet fermions are finely 
degenerate \cite{ks}.   

If we find that there is another possible assignment 
of the lepton number such as $L(\eta)=1$ and $L(N_k)=0$ \cite{ext-s,ks-l},
we can consider a new leptogenesis scenario allowed in this model. 
It is based on the non-thermal generation of lepton number asymmetry 
through the inflaton decay.\footnote{The generation of
lepton number asymmetry through the inflaton decay has been considered.
In \cite{ls}, for example, the asymmetry is supposed to be generated 
by its decay to the heavy right-handed neutrinos and their successive 
decay in the SO(10) GUT framework. Mass of the decay products is 
largely different from the one in the present scenario.}  
Although the quartic coupling $\lambda_5(\eta^\dagger\phi)^2$ 
is forbidden by this lepton number assignment, 
it could be generated effectively through the lepton number 
violating tri-linear scalar couplings $\mu_a$ at low energy 
regions as discussed in the previous section. 
Since this coupling violates the lepton number, 
the decay of inflaton induced through this coupling could generate 
the lepton number asymmetry in the $\eta$ sector through the 
interference between tree and one-loop processes.
The $CP$ asymmetry expected in this decay can be estimated as
\begin{eqnarray}
\epsilon&\equiv&\frac{\Gamma(S_1\rightarrow\eta\phi^\dagger)
-\bar\Gamma(S_1\rightarrow\eta^\dagger\phi)}
{\Gamma(S_1\rightarrow\eta\phi^\dagger)
+\bar\Gamma(S_1\rightarrow\eta^\dagger\phi)} \nonumber \\
&=& \frac{|\mu_2|^2}{8\pi}
\left[\frac{1}{m_{S_1}^2}\ln\frac{(m_{S_1}^2+m_{S_2}^2)}{m_{S_2}^2}  
+ \frac{m_{S_1}^2-m_{S_2}^2}{(m_{S_1}^2-m_{S_2}^2)^2+m_{S_2}^2\Gamma_{S_2}^2}\right]
\sin 2(\theta_1-\theta_2),
\label{cp}
\end{eqnarray}
where $\theta_a={\rm arg}(\mu_a)$ and 
$\Gamma_{S_2}=\frac{1}{8\pi}\frac{|\mu_2|^2}{m_{S_2}}$.
In the following study, we assume the maximum $CP$ phase 
$|\sin 2(\theta_1-\theta_2)|=1$.

Since $\eta_R^0$ is supposed to be DM, all components of $\eta$ 
is expected to be lighter than $N_k$. 
In that case, the generated lepton number asymmetry cannot be 
transferred from the $\eta$ sector to the doublet lepton sector 
through the $\eta$ decay.\footnote{If a singlet fermion is 
considered to be DM, $\eta$ could decay to the lepton 
and then the lepton number asymmetry in the $\eta$ sector 
moves to the lepton sector directly through it \cite{ext-s}. 
In this case, unfortunately, the relic DM abundance could have 
serious tensions with the lepton flavor violating processes 
(LFV) \cite{fcnc}. However, if we assume a certain flavor structure 
for neutrino Yukawa couplings, the LFV constraints could be satisfied.
Detailed analysis for realistic parameters will be presented 
elsewhere.}
However, it can be converted to the lepton sector through 2-2 
scattering processes.
If this conversion occurs efficiently without inducing any contradiction 
with other phenomenological constraints, the sphaleron interaction 
is expected to generate the baryon number asymmetry from 
this lepton number asymmetry. It has already been studied in a different 
inflation scenario \cite{ks-l}.
However, since the parameter space of the present model is much 
simpler than the one of that model, we can study the feature of 
the scenario in a systematic way. 
We start with a brief review for the method of the
analysis at first.
 
The lepton number asymmetry in the co-moving volume is expressed 
by using the entropy density $s$ as 
$\Delta Y_L\equiv\frac{n_\ell-n_{\bar\ell}}{s}$ in the doublet lepton sector 
and as $\Delta Y_\eta\equiv\frac{n_\eta-n_{\eta^\dagger}}{s}$ 
in the $\eta$ sector, respectively. 
Boltzmann equations which describe the evolution of these quantities 
are given as
\begin{eqnarray}
\frac{d\Delta Y_\eta}{dz}&=&-\frac{z}{sH(M_\eta)}\left[
2(\gamma_a+\gamma_b)\left(\frac{\Delta Y_\eta}{Y_\eta^{\rm eq}}-
\frac{\Delta Y_L}{Y_L^{\rm eq}}\right)+
2(\gamma_x+\gamma_y)\frac{\Delta Y_\eta}{Y_\eta^{\rm eq}}\right],\nonumber \\
\frac{d\Delta Y_L}{dz}&=&\frac{z}{sH(M_\eta)}
2(\gamma_a+\gamma_b)\left(\frac{\Delta Y_\eta}{Y_\eta^{\rm eq}}-
\frac{\Delta Y_L}{Y_L^{\rm eq}}\right),
\label{bqn}
\end{eqnarray} 
where we introduce a dimensionless parameter $z$ which 
is defined as $z=\frac{M_\eta}{T}$. 
The equilibrium values for these are expressed as $Y^{\rm eq}_{\eta}(z) = \frac{45}{\pi^4g_*}z^2K_2(z)$ and $Y
^{\rm eq}_L=\frac{81}{\pi^4g_*}$, where $K_2(z)$ is the modified Bessel function of the second kind. 
These equations are derived under the assumption such that the 
existence of $S_{1,2}$ in the thermal bath can be neglected.
It means that both the inverse decay of $S_{1,2}$ and the scattering 
containing $S_{1,2}$ in the initial and final states do not contribute
to these equations.
This requires $T_R<m_{S_{1,2}}$ for the reheating temperature $T_R$. 
If we take account of eqs.~(\ref{parameters}) and (\ref{tr}) with it,  
$\frac{|\mu_1|^2}{m_{S_1}^2}\ll\frac{|\mu_2|^2}{m_{S_2}^2}$ is found 
to be imposed. Reaction densities $\gamma_{a,b}$ 
for lepton number conserving scattering processes
$\eta\eta\rightarrow \ell_\alpha\ell_\beta$ and 
$\eta\ell_\alpha^\dagger\rightarrow\eta^\dagger\ell_\beta$ 
cause the transition of the lepton number asymmetry between the
$\eta$ sector and the doublet lepton sector. 
On the other hand, reaction densities 
$\gamma_{x,y}$ for lepton number violating scattering processes 
$\eta\eta\rightarrow \phi\phi$ and 
$\eta\phi^\dagger\rightarrow \eta^\dagger\phi$ control 
the washout of the lepton number asymmetry in the $\eta$ sector. 
Formulas of these reaction densities are summarized in Appendix A.
If we note that $n_{S_1}(T_R)=\frac{\rho_{S_1}(T_R)}{m_{S_1}}$ and 
$\rho_{S_1}(T_R)=\frac{\pi^2}{30}g_\ast T_R^4$ are satisfied 
for the number density and the energy density of $S_1$ at $T_R$
under the assumption of instantaneous thermalization,
$\Delta Y_\eta(T_R)= \frac{3}{4}\epsilon\frac{T_R}{m_{S_1}}$ could be obtained by using eq.~(\ref{cp}).
We use it and $\Delta Y_L(T_R)=0$ as the initial values for this analysis.
The lepton number asymmetry $\Delta Y_L$ obtained at the weak scale 
as the solution of these Boltzmann equations is converted to 
the baryon number asymmetry through the sphaleron processes. 
In the present model, the resulting baryon number asymmetry 
could be estimated by using the solution of eq.~(\ref{bqn}) 
as\footnote{We find a relation $B=-\frac{7}{19}(B-L)$ 
at the weak scale from the chemical equilibrium condition \cite{ext-s}.
Since $B-L$ is conserved under the sphaleron interaction, we note that 
eq.~(\ref{bqn}) should be regarded as the Boltzmann equations for it 
but not for $L$.} 
\begin{equation}
Y_B=-\frac{7}{19}\Delta Y_L(z_{EW}).
\label{bform}
\end{equation} 

\begin{figure}[t]
\begin{center}
\epsfxsize=7cm
\leavevmode
\epsfbox{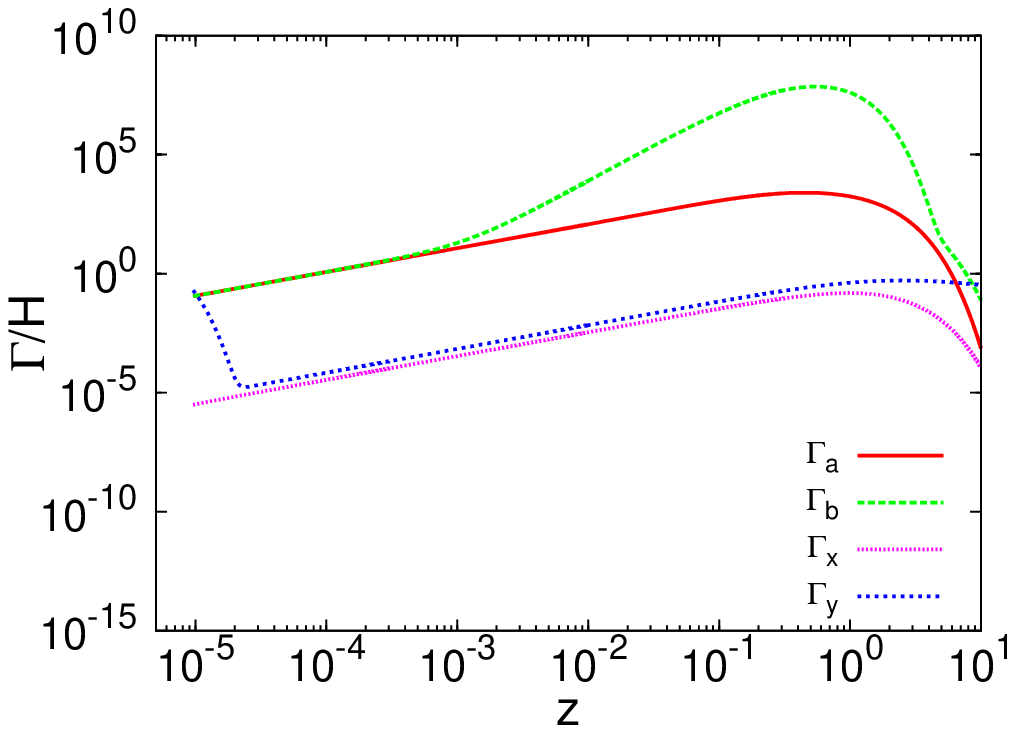}
\hspace*{5mm}
\epsfxsize=7cm
\leavevmode
\epsfbox{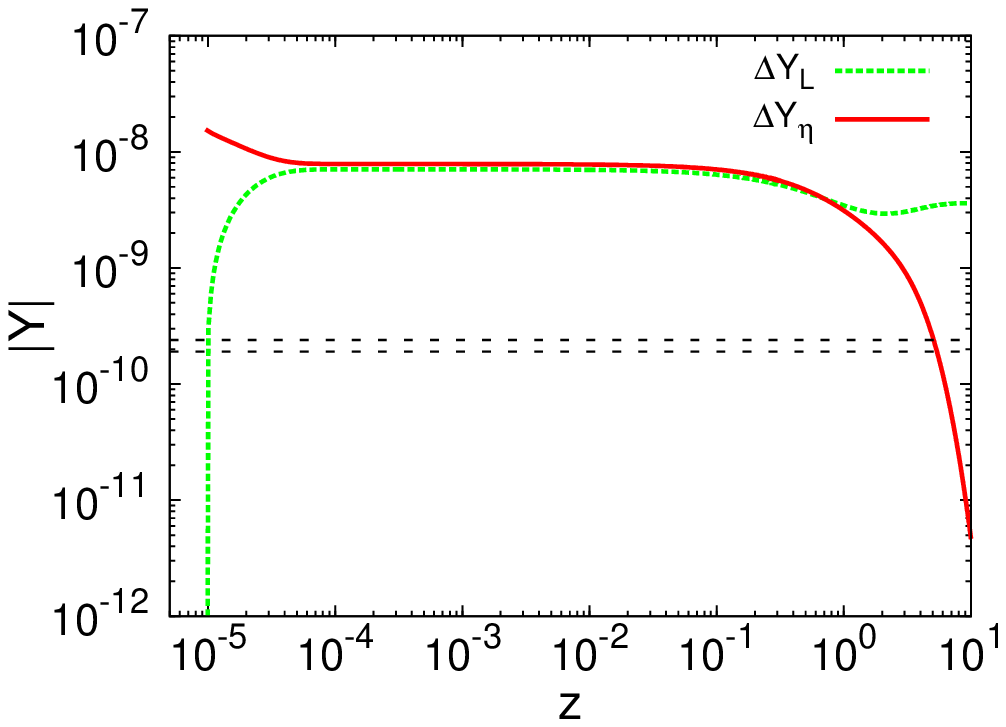}\\
\epsfxsize=7cm
\leavevmode
\epsfbox{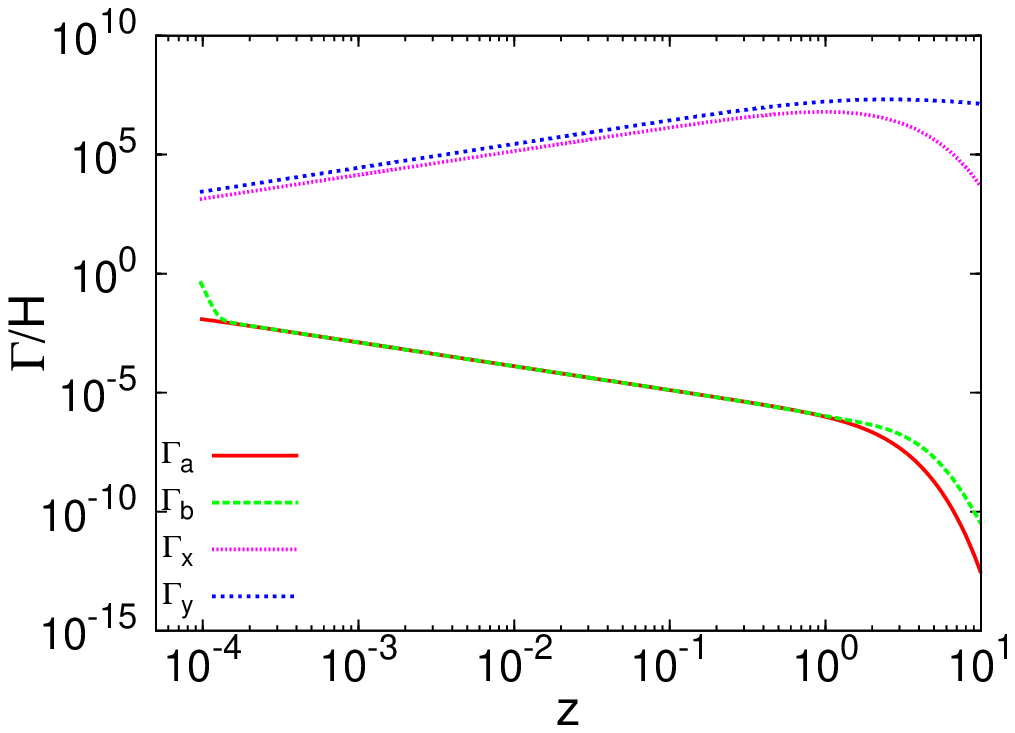}
\hspace*{5mm}
\epsfxsize=7cm
\leavevmode
\epsfbox{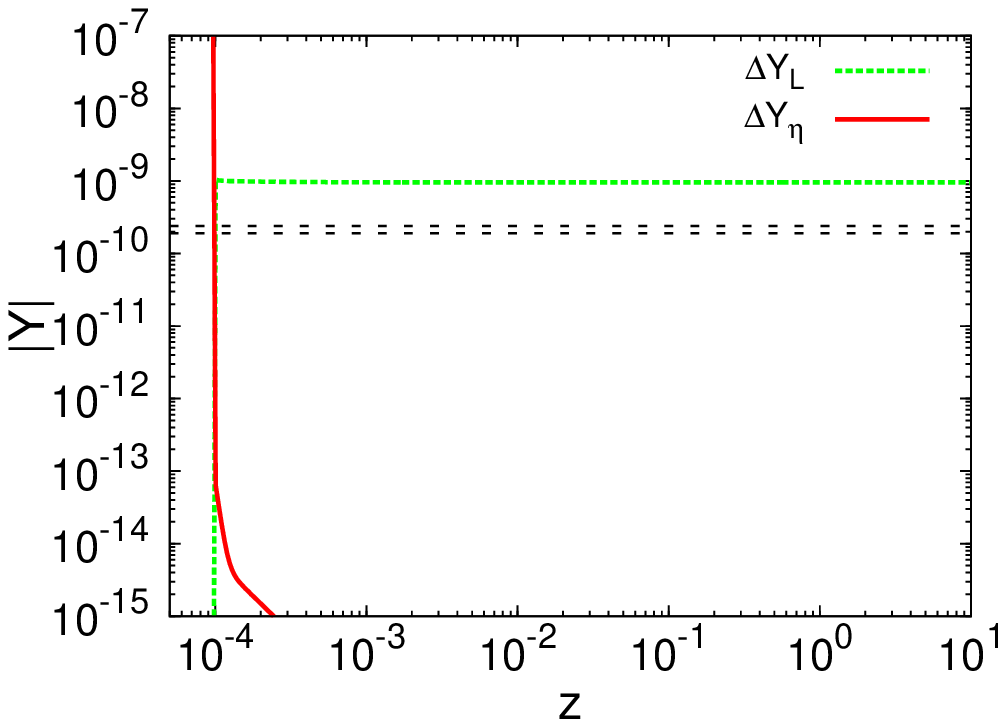} 
\end{center}
\vspace*{-3mm}

{\footnotesize {\bf Fig.~3}~~Relevant reaction rates and 
solutions of the Boltzmann equations for the cases (a) 
(upper panels) and (b) (lower panels).
They correspond to the cases with single fermion masses of
$O(10^4)$ GeV and $O(10^8)$ GeV. 
In the left-hand panels, the ratio of reaction rate $\Gamma$ to the Hubble 
parameter $H$ for the relevant process is plotted as functions of $z$. 
In the right-hand panels, the evolution of $\Delta Y_\eta$ and $\Delta Y_L$ 
are plotted as functions of $z$. 
The lepton number asymmetry required to explain the observational 
results is shown by the horizontal black dashed line.}
\end{figure}

Now we show the results of the numerical analysis of the baryon 
number asymmetry generated through the scenario described above.  
Since a factor $\sum_k(hh^\dagger)_{kk}$ is included in the 
reduced cross section $\hat\sigma_{a,b}$ given in Appendix A, 
we have to determine the flavor structure of neutrino Yukawa couplings 
for the realistic analysis.
We adopt the results obtained in Fig.~2 for it and then they are fixed as
\begin{equation}
(hh^\dagger)_{22}=(1+q_1^2)h_2^2, \qquad (hh^\dagger)_{33}=(1+q_2^2+q_3^2)h_3^2,
\end{equation}
where we could choose the values of $q_{2,3}$ in the allowed region shown 
in Fig.~2 for $q_1=0.77$. In the following analysis, we use
$q_2=-0.4$ and $q_3=2.1$ which fix the neutrino oscillation parameters
as $\Delta m_{32}^2=2.51\cdot 10^{-3}~{\rm eV}^2$, 
$\Delta m_{21}^2=7.63\cdot 10^{-5}~{\rm eV}^2$, 
$\sin^2 2\theta_{23}=0.977$, $\sin^2 2\theta_{12}=0.863$ and 
$\sin^2 2\theta_{13}=0.097$.
In Fig.~3, we show the solutions obtained for the model parameters 
given as the cases (a) and (b) in Table~1.
For each case, the ratio of the relevant reaction rate to the Hubble 
parameter is plotted as a function of $z$ in the left panels.
The evolution of $\Delta Y_{\eta}$ and $\Delta Y_L$ is shown 
in the right panels, where $\Delta Y_L(z_{EW})$ required for 
the observed baryon number asymmetry is implicated by the 
horizontal dotted lines. 
We find that the sufficient amount of baryon number asymmetry is 
generated for parameters which are consistent with the neutrino 
oscillation data. The obtained values are listed in the last column 
of Table~1.

In Fig.~3, we find completely different behavior in the transition of 
the lepton number asymmetry between the two cases. 
The difference comes from whether the lepton number conserving 
scatterings could be in the thermal equilibrium or not.
It is determined depending on both values of the neutrino Yukawa 
couplings and singlet fermion masses.
In this model, they are constrained by the neutrino oscillation data 
together with the value of the effective coupling 
$\lambda_5(\simeq \frac{\mu_2^2}{m_{S_2}^2})$ as found 
in eq.~(\ref{parameters}).
If the lepton number conserving scatterings could be expected in the thermal 
equilibrium for $T_R\gg M_k$, the situation 
$\Delta Y_L\simeq \Delta Y_\eta$ is realized during that period. 
It is found in the case (a).
The final value of $\Delta Y_L$ at the electroweak scale is determined 
depending on the time when they leave the equilibrium.
On the other hand, if we suppose that these processes could never 
be in the thermal equilibrium,  we find that $T_R\ll M_k$ should be 
satisfied and the relation $\Delta Y_L\simeq\Delta Y_\eta$ cannot 
be kept during any substantial period. 
This corresponds to the case (b). 
The final value of $\Delta Y_L$ in this case is fixed mainly 
by the strength of the lepton number conserving scattering 
processes at $T_R$.

\begin{figure}[t]
\begin{center}
\epsfxsize=7.5cm
\leavevmode
\epsfbox{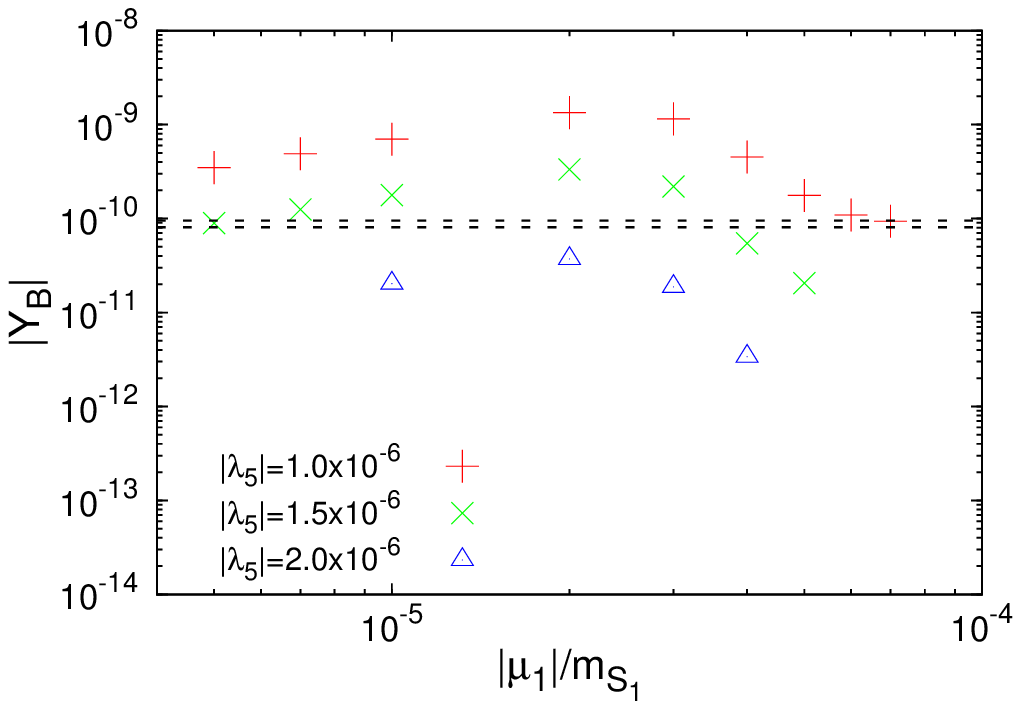} 
\hspace*{5mm}
\epsfxsize=7.5cm
\leavevmode
\epsfbox{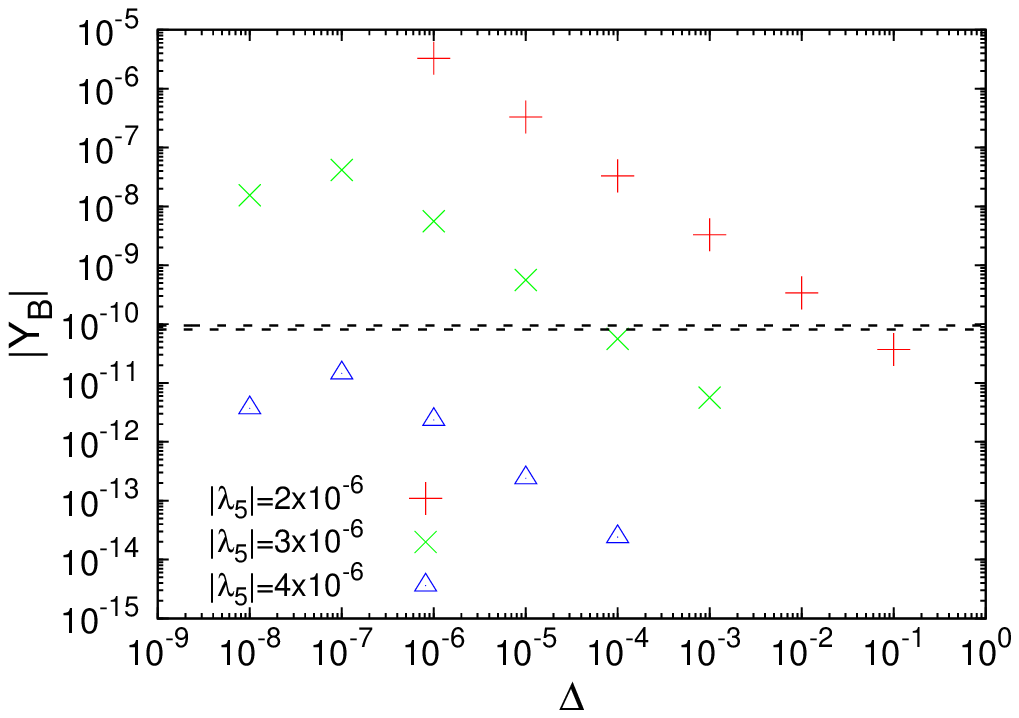}
\end{center}
\vspace*{-3mm}

{\footnotesize {\bf Fig.~4}~~
Left: The dependence of $|Y_B|$ on $\frac{|\mu_1|}{m_{S_1}}$, 
$\frac{|\mu_2|}{m_{S_2}}(\simeq \sqrt{|\lambda_5|})$ which are
relevant to the initial value of the lepton number asymmetry 
and the washout of the lepton number asymmetry, respectively.
$m_{S_2}=1.1m_{S_1}$ is assumed.
Right: The dependence of $|Y_B|$ on 
$\Delta\equiv \frac{m_{S_2}}{m_{S_1}}-1$ for several values of
$|\lambda_5|$. $\frac{|\mu_1|}{m_{S_1}}=2\cdot 10^{-5}$ is assumed.
In both panels, other parameters are fixed to satisfy 
the neutrino oscillation data assuming $m_{S_1}=10^9$~GeV and $M_\eta=1$~TeV 
as discussed in the part relevant to eq.~(\ref{parameters}).}   
\end{figure}

It is useful to clarify the parameter dependence of
the generated baryon number asymmetry $Y_B$ in order to understand 
this scenario.
In the case (a), $Y_B$ is stable against the change of $h_{2,3}$ and 
$M_{2,3}$ which satisfies the conditions given in eq.~(\ref{parameters})
as long as $\lambda_5$ is kept to be a constant value.
This feature is considered to be brought about since the decoupling 
temperature of the lepton number conserving scatterings is not 
affected by this variation substantially.
On the other hand, $Y_B$ is deeply dependent on the values of 
$\mu_1$ and $\mu_2$. The former one determines the initial value 
of $\Delta Y_\eta$ and the latter one controls the washout of 
$\Delta Y_\eta$. 
Examples of this dependence could be seen are through the left panel 
of Fig.~4. 
Smaller $|\mu_1|$ makes the reheating temperature $T_R$ lower and then
reduces the initial lepton number asymmetry $\Delta Y_\eta(T_R)$.
Larger $|\mu_1|$ makes $T_R$ higher and $T_R$ could take a near 
value to $m_{S_{1,2}}$ in the present setting.
This makes the washout of $\Delta Y_\eta$ at the neighborhood of $T_R$
be enhanced due to the tail effect of the $s$-channel resonance.
It can be seen at a small $z$ region in the figures of $\frac{\Gamma}{H}$.
These could explain the reason why $|Y_B|$ is a convex function of
$\frac{|\mu_1|}{m_{S_1}}$. Since larger $|\lambda_5|$ makes the washout
effect larger as found from the formulas of $\hat\sigma_{x,y}$ 
given in Appendix A, $|\lambda_5|$ is expected to have an upper bound.
The left panel of Fig.~4 shows that $|\lambda_5|$ should be smaller than
$2\cdot 10^{-6}$ to guarantee the sufficient amount of $|Y_B|$ in this case. 
Degeneracy between $m_{S_1}$ and $m_{S_2}$ is also crucial to fix the 
value of the $CP$ asymmetry and then the initial lepton number 
asymmetry $\Delta Y_\eta(T_R)$. In the right panel of Fig.~4,
$|Y_B|$ is plotted as a function of $\Delta(\equiv \frac{m_{S_2}}{m_{S_1}}-1)$.
If the mass difference becomes smaller, the second term in eq.~(\ref{cp})
which comes from the self-energy diagram is enhanced and then 
the initial value of $\Delta Y_\eta(T_R)$ becomes larger. 
However, as shown in these examples, the fine degeneracy is not 
required as long as $\lambda_5$ takes a smaller value as mentioned above.
It is a distinctive point from the ordinary TeV scale thermal leptogenesis. 
This comes from the non-thermal origin due to the inflation 
decay which could prepare a sufficient amount of asymmetry.

In the case (b), the situation is completely different from the case (a).
$Y_B$ is largely dependent on the setting of $h_{2,3}$ and $M_{2,3}$ 
for a fixed value of $\lambda_5$.
Unless $M_k$ is much larger than $T_R$, 
the rate of the lepton number conserving scatterings could be enhanced 
to be almost in the thermal equilibrium at the neighborhood of $T_R$.
In that case, since the washout processes are in the equilibrium 
due to an assumed large value of $|\lambda_5|$,
$\Delta Y_\eta$ decreases steeply and the transferred $\Delta Y_L$
follows $\Delta Y_\eta$ not to reach a substantial value. 
To escape this situation, it seems to be necessary 
for $M_k$ to be larger than $T_R$ by an order of magnitude at least.
These features clarify how both $\frac{\mu_1}{m_{S_1}}$ and 
$\frac{\mu_2}{m_{S_2}}$ play crucial roles in this scenario.    
Anyway, these studies suggest that the present 
non-thermal leptogenesis could give us a successful 
scenario for the generation of the baryon number asymmetry.
It is completely consistent with the neutrino mass generation
which could explain all the neutrino oscillation data.

\subsection{Dark matter} 
In this subsection, we discuss the connection between this leptogenesis 
scenario and DM physics under the assumption that the lightest neutral 
scalar $\eta_R^0$ is DM. 
At first, we discuss the relic abundance of $\eta_R^0$.
In the mass range of $\eta_R^0$ discussed in this paper, 
its abundance is known to be determined by the couplings 
$\lambda_{3,4}$ in eq.~(\ref{model}) \cite{ks,ham}, 
which are completely irrelevant to the analysis of other phenomena
studied here.\footnote{We note that the required relic abundance 
cannot be explained for $M_{\eta_R^0}~{^<_\sim}$~530 GeV where only 
the gauge interaction reduces it below the required value \cite{ham}.} 
As discussed in \cite{ks-l}, the lepton number asymmetry kept in 
the $\eta$ sector cannot play any role in the DM abundance.
The asymmetry in this sector disappears 
through the effective coupling $\lambda_5$ after the electroweak symmetry 
breaking. Thus, the required relic abundance should be realized 
as thermal relics for suitable values of $\lambda_{3,4}$.
The estimation of its relic abundance for 
$M_\eta=$~600~GeV, 1 and 3~TeV is shown in Fig.~5, which is obtained 
by using the ordinary method whose detail can be found in \cite{ks-l}. 
Since the required abundance is displayed by the horizontal dotted 
line in this figure, we find that appropriate values of $\lambda_{3,4}$ 
could make the thermal $\eta_R^0$ be a favorable DM candidate naturally.

\begin{figure}[t]
\begin{center}
\epsfxsize=5.1cm
\leavevmode
\epsfbox{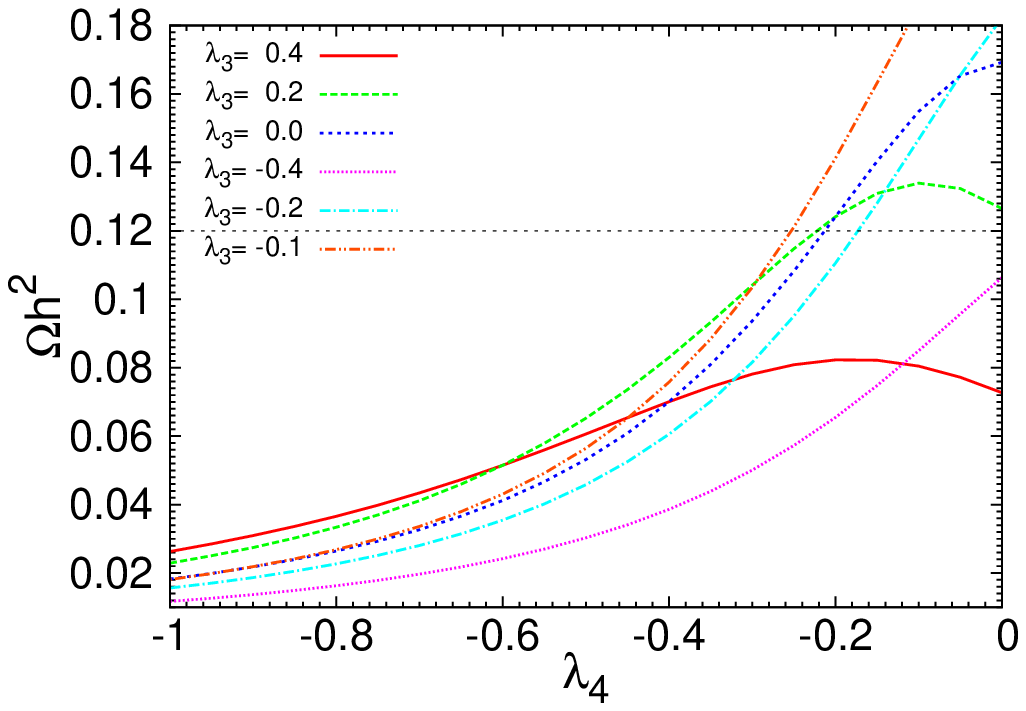}
\hspace*{1mm}
\epsfxsize=5.1cm
\leavevmode
\epsfbox{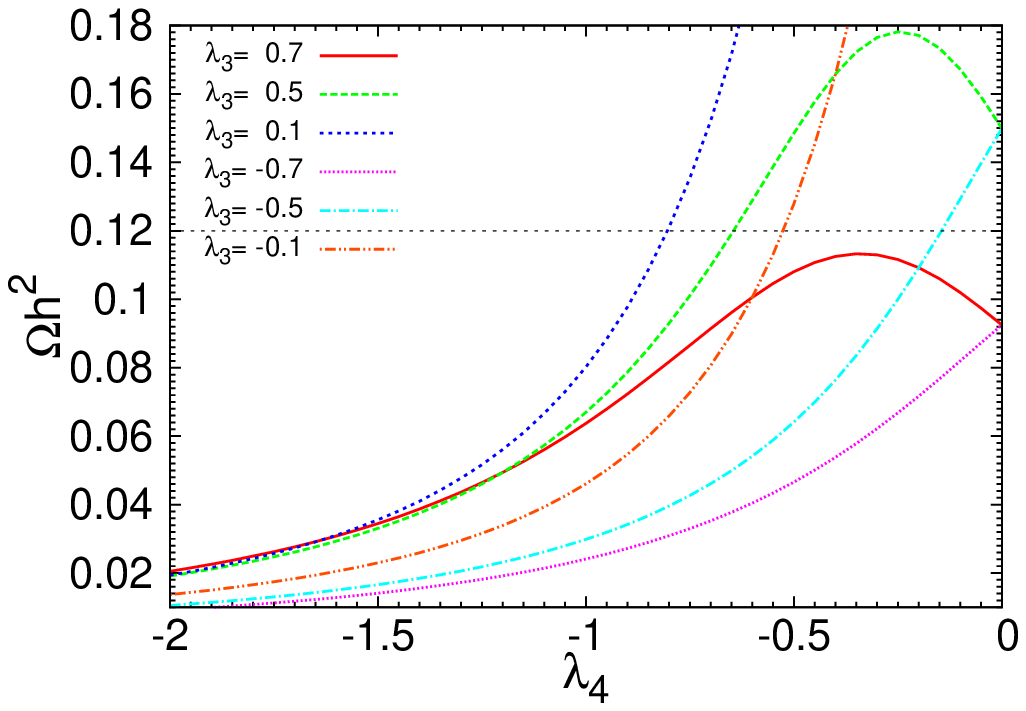}
\hspace*{1mm}
\epsfxsize=5.1cm
\leavevmode
\epsfbox{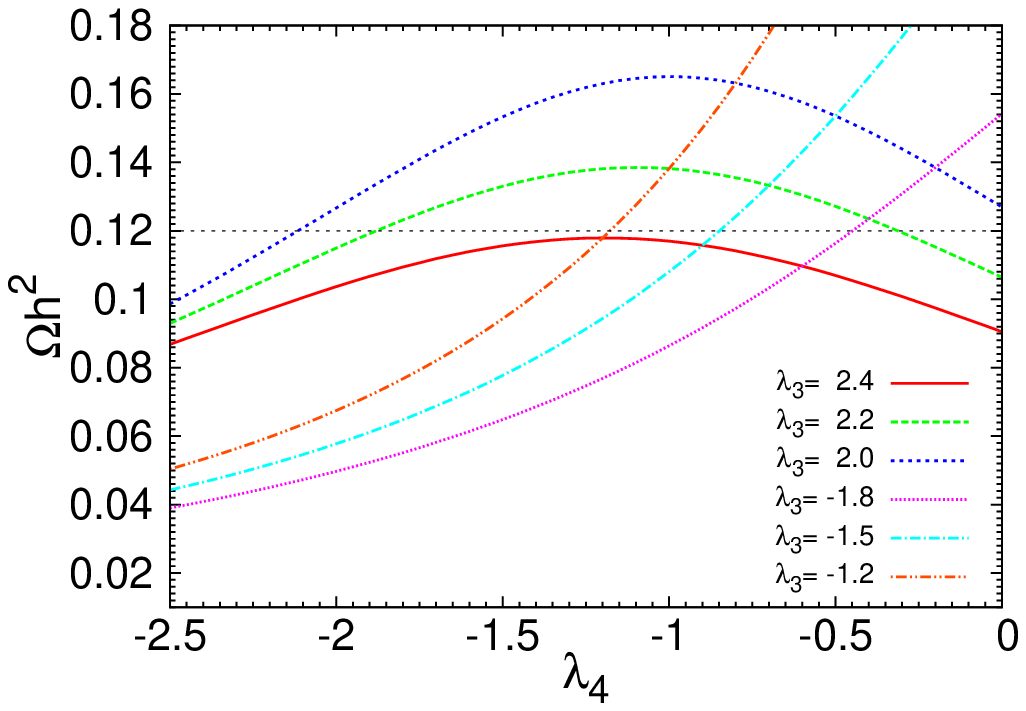}
\end{center}
\vspace*{-3mm}

{\footnotesize {\bf Fig.~5}~~ Thermal $\eta_R^0$ relic abundance 
for $M_\eta=$~600~GeV, 1~TeV and 3~TeV from the left to the right, 
respectively. 
It is estimated by taking account of the coannihilation between 
the components of $\eta$ which is controlled by the couplings 
$\lambda_{3,4}$. }  
\end{figure}

Next, we discuss the consistency between this leptogenesis scenario and
the $\eta_R^0$ DM. Since the effective coupling 
$\lambda_5(\simeq\frac{\mu_2^2}{m_{S_2}^2})$ takes a small value, 
the real and imaginary parts $\eta_{R,I}^0$ of the neutral 
component of $\eta$ have almost degenerate masses whose difference
is expressed as\footnote{ $\eta_R^0$ and $\eta_I^0$ could be mass eigenstates 
for a real $\lambda_5$. Since we assume that $\mu_2$ is real 
and $\frac{|\mu_1|^2}{m_{S_1}^2}\ll\frac{|\mu_2|^2}{m_{S_2}^2}$ 
is satisfied, the effective coupling $\lambda_5$ can be treated as 
a real parameter.}
\begin{equation}
\delta\equiv M_{\eta_I^0}-M_{\eta_R^0}=
\frac{\langle\phi\rangle^2}{M_\eta}\frac{\mu_2^2}{m_{S_2}^2}.
\end{equation}
On the other hand, $\eta_R^0$ and $\eta_I^0$ have a weak
gauge interaction such as
\begin{equation}
{\cal L}_{\rm int}= \frac{g}{2}Z^\mu
(\eta_R\partial_\mu\eta_I-\eta_I\partial_\mu\eta_R).
\end{equation}
This interaction induces the spin-independent 
inelastic scattering $\eta_R N\rightarrow \eta_IN$ with a nucleus $N$
mediated by a $Z^0$ exchange in which the nucleus is not excited 
as long as the mass difference $\delta$ is sufficiently small. 
This reaction can contribute to the direct search of DM. 
The $\eta_R^0$-nucleon cross section for this inelastic scattering can be 
estimated as
\begin{equation}
\sigma_{n,{\rm inel}}^0=\frac{G_F^2}{2\pi\mu_n^2}
\simeq 7.44\times 10^{-39}~{\rm cm}^2,
\label{inelcross}
\end{equation}
where $\mu_n$ is the reduced mass of this $\eta_R^0$-nucleon system
and $\delta\ll M_\eta$ is assumed.
If we apply the bound from recent DM direct search experiments
to this cross section, we could approximately derive 
a useful constraint on this leptogenesis scenario.

\begin{figure}[t]
\begin{center}
\epsfxsize=8cm
\leavevmode
\epsfbox{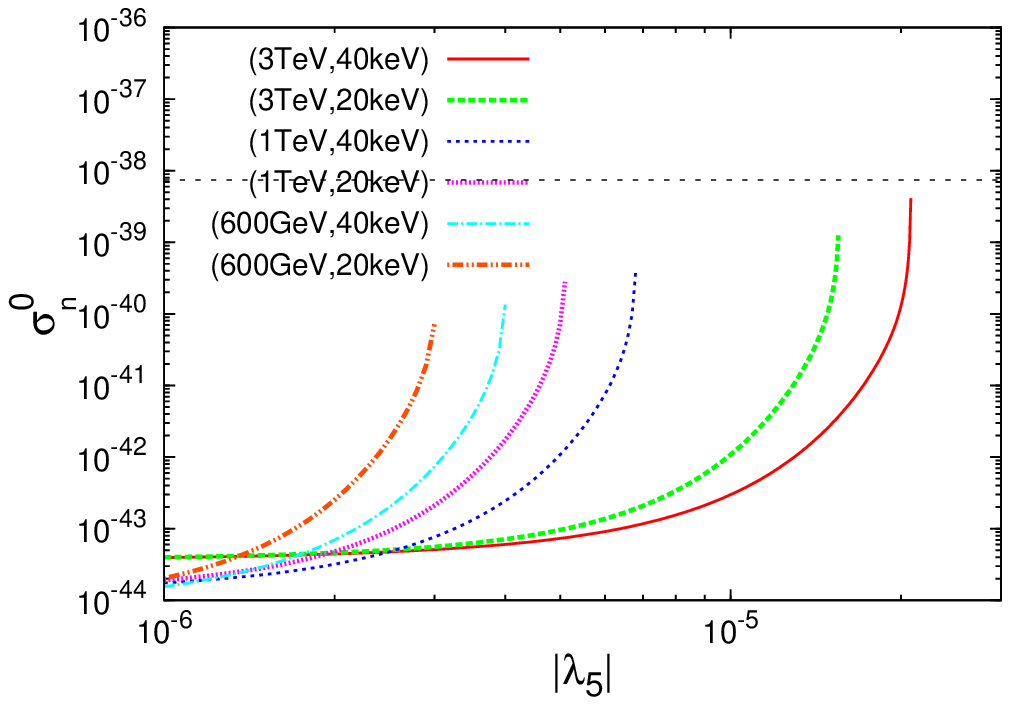}
\end{center}
\vspace*{-3mm}

{\footnotesize {\bf Fig.~6}~~Constraint on the effective 
coupling $\lambda_5$, which is derived from the DM direct detection 
experiments. Each line represents $\sigma_{n,{\rm inel}}^0$ for $(M_\eta,E_R)$.
A bound for $\sigma_n^0$ at $m_{\rm DM}$= 3~TeV, 1~TeV and 
600~GeV given by LUX is used. }  
\end{figure}

As briefly described the in Appendix B, the present bound on $\sigma_n^0$
estimated for the elastic scattering $(\delta=0)$ might be 
translated to the one for the inelastic scattering $(\delta\not=0)$ as 
\begin{equation}
\sigma_{n,{\rm inel}}^0 =\sigma_{n,{\rm el}}^0\frac{
\displaystyle\int_{v_{\rm min}(\delta=0)}^{v_{\rm esc}} dv
\left(e^{-\frac{(v-v_e)^2}{v_0^2}}
-e^{-\frac{(v+v_e)^2}{v_0^2}}\right)}
{\displaystyle \int_{v_{\rm min}(\delta\not=0)}^{v_{\rm esc}} dv
\left(e^{-\frac{(v-v_e)^2}{v_0^2}}-e^{-\frac{(v+v_e)^2}{v_0^2}}\right)},
\label{inel} 
\end{equation}
where $v_{\rm min}$ is the minimum $\eta_R^0$ velocity required to induce 
this scattering. 
This $\sigma_{n,{\rm inel}}^0$ obtained by using the present bound on 
$\sigma_n^0$ should be larger than the value given in eq.~(\ref{inelcross}) 
since DM has not been detected in any direct detection experiments 
\cite{direct1,direct2}.
We also note that $\eta_R^0$ cannot be detected in any direct 
search experiments unless $v_{\rm min} < v_{\rm esc}$ is satisfied
for the local escape velocity $v_{\rm esc}$ from our Galaxy. 
It is estimated as 498 km/s$<v_{\rm esc}<$608 km/s \cite{esc} and 
its medium value 544 km/s is used as $v_{\rm esc}$ in this analysis. 
In Fig.~6, we plot $\sigma_{n,{\rm inel}}^0$ corresponding 
to it as a function of $|\lambda_5|$ for typical values of 
the recoil energy of nucleus $E_R$ by using the bound of $\sigma_n^0$ 
given by LUX \cite{direct2} as $\sigma_{n,{\rm el}}^0$ 
in eq.~(\ref{inel}).\footnote{Since the differential WIMP-nucleus cross section in the case of scattering
leading to a heavier WIMP depends on the recoil energy very differently from
that associated with the standard elastic scattering and since in the present
case the nucleon cross section depends on the WIMP velocity, clearly we cannot
derive an exact bound by using the experimental result for the elastic scattering.}
Since the endpoint of each line represents the occurrence 
of $v_{\rm min}=v_{\rm esc}$, the scattering is kinematically 
forbidden at the larger $|\lambda_5|$ region than it 
and then such a region of $|\lambda_5|$ is allowed from the result 
of the present direct DM search.
Unless the escape velocity takes much larger values than the one used here, 
this analysis unfortunately suggests that 
this DM with the mass $M_\eta ~{^<_\sim}~6$~TeV is difficult to 
be detected through this inelastic scattering process in the 
direct search experiments.

If we take account of the relation 
$|\lambda_5|\simeq \frac{|\mu_2|^2}{m_{S_2}}$ in this model, 
we might roughly read off the condition required from this figure 
as follows,
\begin{equation}
\frac{\mu_2^2}{m_{S_2}^2}~{^>_\sim}~\left\{
\begin{array}{ll} 
(1-2)\times 10^{-5}  & \qquad{\rm for}\ M_{\eta_R^0}=3~{\rm TeV}, \\
(5-7)\times 10^{-6}  & \qquad{\rm for}\ M_{\eta_R^0}=1~{\rm TeV}, \\
(3-4)\times 10^{-6}  & \qquad{\rm for}\ M_{\eta_R^0}=600~{\rm GeV}, \\
\end{array} \right.
\label{clam5}
\end{equation}
although it has non-negligible dependence on $E_R$ and $v_{\rm esc}$.
We should note that $|\lambda_5|$ cannot take small 
values freely even if the neutrino mass constraints have only to be satisfied.
In the case (b), this constraint is found to be consistent with 
the parameters used in the above analysis of the baryon number 
asymmetry.
On the other hand, in the case (a), one might consider from Fig.~6
that this constraint is not satisfied and the scenario is inconsistent. 
However, it may be appropriate to judge that the consistency of 
the parameters used here is marginal if we take account of several 
uncertainties included in the estimation of the bound on $|\lambda_5|$.
We also note that the situation could be improved if we suppose 
fine mass degeneracy between $S_1$ and $S_2$ such as $\Delta=10^{-6}$ 
and a favorable value for $\frac{|\mu_1|}{m_{S_1}}$ such as $2\cdot 10^{-5}$. 
As we find in the right-panel of Fig.~5, these could enhance the 
initial asymmetry of the lepton number produced through the inflaton decay.
In this case, for example, $|Y_B|=1.2\cdot 10^{-10}$ can be obtained 
for $|\lambda_5|=3.5\cdot 10^{-6}$ at $M_\eta=1$~TeV.

\section{Summary}
We have proposed an extension of the radiative neutrino mass model 
with $Z_2$ odd real singlet scalars, which give a seed of lepton number 
violation to allow Majorana neutrino mass generation at one-loop level.
If they have hierarchical non-minimal couplings with the Ricci scalar
such as $\sum_a\xi_aS_a^2R$ with $\xi_1\gg 1\gg\xi_2$, 
the $S_1$ potential at large field regions 
is so flat that sufficient inflation could be induced. 
Although both the scalar spectral index and the tensor-to-scalar ratio 
take favorable values just as the Higgs inflation, the model can evade
from the unitarity problem differently from the ordinary Higgs inflation.
Since the unitarity violating scale could be similar to or larger than 
the inflation scale, the flatness of the inflaton potential is never 
disturbed by new physics which restores the unitarity of the model. 

The decay of inflaton could non-thermally produce the lepton number 
asymmetry in the $\eta$ sector through the reheating processes. 
Although its decay cannot yield this asymmetry directly 
in the lepton sector, the lepton number conserving scatterings 
could convert a sufficient amount of asymmetry from the $\eta$ 
sector to the lepton sector.
This lepton number asymmetry is transferred to the baryon number 
asymmetry through the sphaleron interaction.
Parameters relevant to this leptogenesis are constrained by 
the neutrino oscillation data and the DM direct search experiments.    
We have shown that the sufficient baryon number could be generated 
consistently with these constraints under suitable conditions. 
Moreover, the DM abundance could 
be fixed by the parameters which are irrelevant to all of these 
as long as $\eta_R^0$ is supposed to be DM.
Although the model considered here is very simple, it can compactly
explain problems in the SM such as the neutrino mass generation, 
the DM origin and its abundance, the inflation and 
the baryon number asymmetry in the Universe. They are closely related
each other through the radiative mechanism for the neutrino mass generation.

\section*{Acknowledgements}
R.~H.~S.~Budhi is supported by the Directorate General of Higher 
Education (DGHE) of Indonesia (Grant Number 1245/E4.4/K/2012). 
S.~K. is supported by Grant-in-Aid for JSPS fellows (26$\cdot$5862).
D.~S. is supported by MEXT Grant-in-Aid for Scientific Research 
on Innovative Areas (Grant Number 26104009).

\newpage
\section*{Appendix A}
We summarize the formulas for the reaction densities which
contribute to the Boltzmann equations used in this analysis.
We introduce dimensionless variables
\begin{equation}
x=\frac{s}{M_\eta^2}, \qquad a_k=\frac{M_k^2}{M_\eta^2}, \qquad 
b_a=\frac{m_{S_a}^2}{M_\eta^2}, \qquad 
c_a=\frac{|\mu_a|^2}{M_\eta^2},
\end{equation}
where $s$ is the squared center of mass energy.
The reaction density for the scattering processes is expressed as
\begin{equation}
\gamma(ab\rightarrow ij)=\frac{T}{64\pi^4}\int^\infty_{s_{\rm min}}ds~
\hat\sigma(s)\sqrt{s}K_1\left(\frac{\sqrt{s}}{T}\right),
\end{equation}
where $\hat\sigma(s)$ is the reduced cross section and
$K_1(z)$ is the modified Bessel function of the second kind.
The lower bound of integration is defined as 
$s_{\rm min}={\rm max}[(m_a+m_b)^2,(m_i+m_j)^2]$.  

The reduced cross section $\hat\sigma_{a,b}$ for the lepton number 
conserving scattering processes are given as 
\begin{eqnarray}
\hat\sigma_a(x)&=&\frac{1}{2\pi}
\left[\sum_{k=1}^3(hh^\dagger)^2_{kk}\left\{
\frac{a_k(x^2-4x)^{1/2}}{a_kx+(a_k-1)^2}\right.\right. \nonumber \\
&+&\left.\left.
\frac{a_k}{x+2a_k-2}
\ln\left(\frac{x+(x^2-4x)^{1/2}+2a_k-2}
{x-(x^2-4x)^{1/2}+2a_k-2}\right)\right\}
\right.\nonumber\\
&+&\left.
\sum_{i>j}
\frac{{\rm Re}[(hh^\dagger)_{ij}^2]\sqrt{a_ia_j}}{x+a_i+a_j-2}
\left\{
\frac{2x+3a_i+a_j-4}{a_j-a_i}
\ln\left(\frac{x+(x^2-4x)^{1/2}+2a_i-2}
{x-(x^2-4x)^{1/2}+2a_i-2}\right)\right.\right. \nonumber \\
&+&\left.\left. \frac{2x+a_i+3a_j-4}{a_i-a_j}
\ln\left(\frac{x+(x^2-4x)^{1/2}+2a_j-2}
{x-(x^2-4x)^{1/2}+2a_j-2}\right)
\right\}\right] 
\label{lv2}
\end{eqnarray}
for $\eta\eta \rightarrow \ell_\alpha\ell_\beta $ and 
\begin{eqnarray}
\hat\sigma_b(x)&=&\frac{1}{2\pi}\frac{(x-1)^2}{x^2}
\left[\sum_{k=1}^3(hh^\dagger)_{kk}^2\frac{a_k}{x}
\left\{\frac{x^2}{xa_k -1}+\frac{2x}{D_k(x)}
+\frac{(x-1)^2}{2D_k(x)^2}\right.\right.\nonumber \\
&-&\left.\frac{x^2}{(x-1)^2}
\left(1+\frac{2(x+a_k-2)}{D_k(x)}\right)
\ln\left(\frac{x(x+a_k-2)}{xa_k-1}\right)\right\}\nonumber \\
&+&\left.
\sum_{i>j}{\rm Re}[(hh^\dagger)_{ij}^2]\frac{\sqrt{a_ia_j}}{x}\left\{
\frac{x}{D_i(x)}+\frac{x}{D_j(x)}+\frac{(x-1)^2}{D_i(x)D_j(x)}
\right.\right.\nonumber \\
&+&\left.\left.\frac{x^2}{(x-1)^2}
\left(\frac{2(x+a_i-2)}{a_j-a_i}-
\frac{x+a_i-2}{D_j(x)}\right)\ln\frac{x(x+a_i-2)}{xa_i-1}
\right.\right. \nonumber\\
&+&\left.\left.\frac{x^2}{(x-1)^2}
\left(\frac{2(x+a_j-2)}{a_i-a_j}-
\frac{x+a_j-2}{D_i(x)}\right)\ln\frac{x(x+a_j-2)}{xa_j-1}
\right\}\right]
\label{lv1}
\end{eqnarray}
for $\eta\ell_\alpha^\dagger \rightarrow \eta^\dagger\ell_\beta$.
In these formulas, we use the following definition for convenience:
\begin{equation}
\frac{1}{D_k(x)}=\frac{x-a_k}{(x-a_k)^2+a_k^2d_k}, \qquad 
d_k=\frac{1}{64\pi^2}\left(\sum_{\alpha=e,\mu,\tau}
|h_{\alpha k}|^2\right)^2\left(1-\frac{1}{a_k}\right)^4.
\end{equation}

If we take account of the assumption 
$\frac{|\mu_1|^2}{m_{S_1}^2}\ll\frac{|\mu_2|^2}{m_{S_2}^2}$,
the reduced cross section $\hat\sigma_{x,y}$ of the lepton number violating 
scattering processes could be approximately represented as 
\begin{eqnarray}
\hat\sigma_x(x)&\simeq&\frac{c_2^2}{\pi}
\frac{1}{(x^3(x-4))^{1/2}}
\left(\frac{2}{P_2^2-1}+ \frac{1}{P_2}
\ln\frac{P_2+1}{P_2-1} \right), \nonumber \\
\hat\sigma_y(x)&\simeq&\frac{c_2^2}{\pi}\left[
\frac{2}{(x-1)^2}\frac{1}{Q_2^2-1} 
+\frac{(x-1)^2}{2x^2}\frac{1}{\tilde D_2(x)}  
+\frac{1}{x} \frac{x-b_2}{\tilde D_2(x)}
\ln\frac{Q_2 +1}{Q_2 -1}\right]
\end{eqnarray}
for $\eta\eta\rightarrow\phi\phi$ and 
$\eta\phi^\dagger \rightarrow\eta^\dagger\phi$, respectively.
In these formulas we use the definition such as
\begin{eqnarray}
&&\frac{1}{\tilde D_a(x)}=\frac{1}
{(x-b_a)^2+b_a^2\tilde d_a}, \qquad
\tilde d_a=\frac{1}{64\pi^2}
\left(\frac{c_a}{b_a}\right)^2\left(1-\frac{1}{b_a}\right)^2, \nonumber \\ 
&&P_a=\frac{2(1-b_a)-x}{[x(x-4)]^{1/2}}, \qquad
Q_a=-1 +\frac{2(1-xb_a)}{(x-1)^2}.
\end{eqnarray}

\section*{Appendix B}
We consider the direct DM detection through the inelastic scattering with 
the target composed of the nucleus with the atomic number $Z$ 
and the mass number $A$ \cite{l5,inel}.
Its differential detection rate per unit target mass is
\begin{equation}
\frac{dR}{dE_R}=N_T\frac{\rho_{\rm DM}}{m_{\rm DM}}
\int d^3v~ v f(\vec{v},\vec{v_e})\frac{d\sigma}{dE_R},
\label{drate}
\end{equation}
where $N_T$ is a number of target nuclei per unit mass and $E_R$ is the
recoil energy of nucleus.
The DM velocity distribution in the rest frame of detector may 
be taken as a Maxwell-Boltzmann distribution 
$f(\vec{v},\vec{v}_e)=
\frac{1}{(\pi v_0^2)^{3/2}}\exp\left(-\frac{(\vec{v}+\vec{v}_e)^2}{v_0^2}\right)$
with $v_0=220$~km/s. We take into account the motion of the Sun and 
the Earth by using $\vec{v}_e$ whose magnitude changes as 
$v_e=v_0\left[1.05+0.07\cos\left(\frac{2\pi(t-t_p)}{1~{\rm yr}}\right)\right]$.
The minimum velocity for which the scattering can occur is estimated as
\begin{equation}
v_{\rm min}=\frac{1}{\sqrt{2m_NE_R}}\left(\frac{m_NE_R}{m_r} +\delta\right),
\end{equation}
where $m_N$ is the mass of a target nucleus and $m_r$ is the reduced 
mass of DM-nucleus system $m_r=\frac{m_{\rm DM}m_N}{m_{\rm DM}+m_N}$.
The differential cross section $\frac{d\sigma}{dE_R}$ for spin independent
interaction is expressed by using the DM-nucleon cross section $\sigma_n^0$ 
at zero momentum transfer as
\begin{equation}
\frac{d\sigma}{dE_R}=\frac{m_N}{2v^2\mu_n^2}\frac{[Zf_p+(A-Z)f_n]^2}{f_n^2}
\sigma_n^0F^2(E_R),
\label{dcross}
\end{equation}
where $F(E_R)$ is a form factor of the nucleus. 
If we substitute this in eq.~(\ref{drate}), the differential rate 
can be represented by using eq.~(\ref{dcross}) as 
\begin{equation}
\frac{dR}{dE_R}=D_N\sigma_n^0\int d^3v~\frac{1}{v}~f(\vec{v},\vec{v_e}),
\end{equation}
where $D_N$ is written as
\begin{equation}
D_N=N_T\frac{m_N\rho_{\rm DM}}{2\mu_n^2m_{\rm DM}}\frac{[Zf_p+(A-Z)f_n]^2}{f_n^2}
F^2(E_R).
\end{equation}
We note that $D_N$ takes a fixed value as long as the same target is used. 

The remaining part depends on the sub-process and
it can be expressed as
\begin{eqnarray}
\sigma_n^0\int dv ~\frac{1}{v^2}~f(\vec{v},\vec{v}_e)=\sigma_n^0
\frac{1}{\sqrt\pi v_0v_e}\int_{v_{\rm min}}^{v_{\rm esc}}dv\left(
e^{-\frac{(v-v_e)^2}{v_0^2}}-e^{-\frac{(v+v_e)^2}{v_0^2}}\right).
\end{eqnarray}
This depends on whether the scattering occurs 
elastically $(\delta=0)$ or inelastically $(\delta\not=0)$.
If we interpret the present direct detection results based on these
scattering processes, the present bound on the elastic scattering 
cross section $\sigma_{n,{\rm el}}^0$ can be translated to the bound 
on the inelastic scattering cross section $\sigma_{n,{\rm inel}}^0$ through
\begin{equation}
\sigma_{n,{\rm inel}}^0\int_{v_{\rm min}(\delta\not=0)}^{v_{\rm esc}}dv
\left(e^{-\frac{(v-v_e)^2}{v_0^2}}-e^{-\frac{(v+v_e)^2}{v_0^2}}\right) 
=\sigma_{n,{\rm el}}^0\int_{v_{\rm min}(\delta=0)}^{v_{\rm esc}}dv
\left(e^{-\frac{(v-v_e)^2}{v_0^2}}-e^{-\frac{(v+v_e)^2}{v_0^2}}\right).
\end{equation}
We use this relation to constrain the allowed values of $\delta$,
which makes us possible to find the lower bound of 
the effective coupling $|\lambda_5|$.

\newpage
\bibliographystyle{unsrt}

\end{document}